\documentclass[10pt]{article}

\usepackage{amsmath}
\usepackage{graphicx}
\usepackage{indentfirst}
\usepackage{amssymb}
\usepackage{cite}
\usepackage{color}
\usepackage{subfigure}

\begin{document}

\title{\bf{Weak Cosmic Censorship Conjecture\\in Myers-Perry Black Hole with Separability}}

\date{}
\maketitle

\begin{center}
\author{Bogeun Gwak}$^a$\footnote{rasenis@dgu.ac.kr}\\

\vskip 0.25in
$^{a}$\it{Division of Physics and Semiconductor Science, Dongguk University, Seoul 04620,\\Republic of Korea}\\
\end{center}
\vskip 0.6in

{\abstract
{We investigate the weak cosmic censorship conjecture in Myers-Perry black holes with arbitrary rotations in general dimensions based on the scattering of a massless scalar field. From the fluxes of the scalar field flowing into the black hole, the changes in mass and angular momenta of the black hole are obtained. However, the extremal and near-extremal black holes with the aforementioned changes are still black holes in the final state. Hence, the conjecture is valid for our investigation. Furthermore, we analyze the changes in the black hole from a thermodynamic perspective to highlight that the laws of thermodynamics support the conjecture.}}

\thispagestyle{empty}
\newpage
\setcounter{page}{1}

\section{Introduction}\label{sec:01}

Black holes are one of the most interesting compact objects in the universe. One of the features of black holes is singularity, which is the endpoint of the geodesics passing through an event horizon. This implies that no geodesic can reach an outside observer, and thus no observer can see the internal structure of the black holes. However, by including quantum processes in the vicinity of black holes, we realized that black holes can emit energy called Hawking radiation\cite{Hawking:1975vcx,Hawking:1976de}. Based on this, the Hawking temperature of the black holes is defined as thermal temperature. From a thermodynamic perspective, the thermodynamic conjugate of temperature is entropy. Black holes also have an entropy associated with an irreducible mass of a black hole calculated by a classical process\cite{Christodoulou:1970wf,Bardeen:1970zz,Christodoulou:1971pcn}. Physically, the irreducible mass is the energy distributed on the horizon\cite{Smarr:1972kt}, and summing it with reducible energies, such as rotational and electric energies, yields the total mass. Because the behavior of irreducible mass is similar to that of entropy, the Bekenstein-Hawking entropy of black holes is defined to be proportional to the square of irreducible mass\cite{Bekenstein:1973ur,Bekenstein:1974ax}. This implies that the surface area of the event horizon is proportional to entropy and is irreducible. Thermodynamic variables can then be defined based on the parameters of a black hole. Furthermore, the laws of thermodynamics including the third law\cite{Bardeen:1973gs} can be computed for a black hole using these variables.

The static observer cannot see the inside of black holes because of the event horizon. In particular, singularity is the endpoint of geodesics inside black holes. This is a problematic point in the causal structure of spacetime. Here, the event horizon plays a crucial role in preventing the observation of singularity. However, the horizon does not always surround the singularity, and its stability depends on the black hole parameters. Some black holes have extremal conditions that exist in the horizon. However, if the extremal condition is violated, spacetime will not have a horizon; therefore, the problematic point is exposed to the observer. This is known as naked singularity. To ensure completeness of the black hole spacetime for a static observer, the horizon should stably surround the singularity. This is known as weak cosmic censorship (WCC) conjecture\cite{Penrose:1964wq,Penrose:1969pc}. Interestingly, WCC conjecture has no general proof of its application to any black hole. Hence, an investigation of WCC conjecture is essential for each black hole. The first test was performed on the extremal Kerr black hole\cite{Wald:1974ge} by adding a particle. It was shown that the conjecture is valid. Actual tests showed that the conclusion depended on the state of the black hole and methods used. For instance, the near-extremal Kerr black hole violated the conjecture by adding a particle\cite{Jacobson:2009kt}; however, this also produced the opposite conclusion when considering self-force\cite{Barausse:2010ka,Colleoni:2015ena,Sorce:2017dst}. A similar situation was observed in Reissner-Nordstr\"{o}m black hole\cite{Hubeny:1998ga,Isoyama:2011ea}. WCC conjecture is investigated by adding a particle on various black holes and their prospects\cite{Bouhmadi-Lopez:2010yjy,Gwak:2011rp,Rocha:2011wp,Gao:2012ca,Hod:2013oxa,Rocha:2014jma,Gwak:2015fsa,Cardoso:2015xtj,Siahaan:2015ljs,Horowitz:2016ezu,Revelar:2017sem,Song:2017mdx,Yu:2018eqq,Gwak:2018tmy,Mcinnes:2019dcw,Zeng:2019huf,Wang:2019jzz,He:2019kws,Hu:2019zxr,Wang:2020osg,Shaymatov:2020wtj,Ying:2020bch,Ahmed:2020ksf,Khodabakhshi:2020fkd,McInnes:2021zlt,Liang:2021xny,Qu:2021hxh}. On the other hand, test fields were also considered instead of a particle. Then, the tests were elevated to the scattering problems of black holes and other related interesting phenomena, such as perturbation, superradiance, and quasinormal modes. Furthermore, WCC conjecture has been studied from various perspectives and distinct approaches\cite{Hod:2008zza,Semiz:2005gs,Toth:2012vvy,Semiz:2015pna,Natario:2016bay,Duztas:2017lxk,Gwak:2018akg,Chen:2019nsr,Natario:2019iex,Gwak:2019asi,Jiang:2019vww,Wang:2019bml,Gwak:2019rcz,Yang:2020iat,Hong:2020zcf,Feng:2020tyc,Yang:2020czk,Gwak:2020zht,Goncalves:2020ccm,Duztas:2021kuj,Gwak:2021tcl,Shaymatov:2022ako,Yang:2022yvq}. Because extremal matter interacts with black hole as a perturbation in the WCC conjecture test, the test should include changes in the black hole such that the changes are closely associated with thermodynamics. In particular, the laws of thermodynamics are consistent with these changes and play an important role in ensuring the validity of WCC conjecture. The first law shows the dispersion relation between variables that vary due to carried conserved charges; similarly, the second law is also satisfied according to the WCC conjecture test. Furthermore, it was observed that the third law is consistent with WCC conjecture because the nonextremal black hole cannot be extremal by a finite number of physical processes. This implies that the quantum physical process represented by thermodynamics is closely related to WCC conjecture from a classical point of view.

The equations of motion for the particles and fields can be divided into individual equations using separation constants. This is an important property for solving the equations, and the separation constants can be obtained from the isometries of spacetime. In spherical symmetric black holes, the separation constants correspond to the conserved charges even if spacetime is at higher dimensions. However, to separate the equations of motion in Kerr black holes, an additional constant must be introduced. In contrast to the separation constants from isometries, the additional constant originates from a hidden symmetry related to an irreducible
rank-two Killing tensor\cite{Carter:1968rr,Carter:1968ks}. The separability of the equations of motion depends on the black holes and matter in the background; therefore, each case has been studied separately for Kerr black holes and other black holes\cite{Carter:1968rr,Carter:1968ks,Teukolsky:1972my,Chandrasekhar:1976ap,Teukolsky:1973ha,Krtous:2018bvk,Vasudevan:2004ca,Krtous:2007xf,Frolov:2006dqt,Frolov:2008jr,Oota:2007vx,Lunin:2017drx}. Based on the hidden symmetries to spacetime, the equations of motion are solvable and understandable; therefore, this is an important property for studying the dynamics of test particles and fields. 

In this work, we investigated the WCC conjecture for Myers-Perry (MP) black hole based on the scattering of a massless scalar field. The MP black hole is a higher-dimensional generalization of the Kerr black hole. Therefore, it can have multiple arbitrary rotations. This is an interesting property because many independent angular momenta that form a new balance with higher-dimensional gravitational force can be developed. However, the metric of the MP black hole with arbitrary rotations is complicated, and the field equations are difficult to solve without separability. This is similar to the problem faced during the investigation of WCC conjecture. For instance, without full separation for arbitrary rotational angles, we can prove the validity of WCC conjecture in limited cases\cite{Gwak:2011rp} by adding a particle. For fixed dimensions, investigations of WCC conjecture are also one way\cite{Shaymatov:2019pmn}. Recently, WCC conjecture has been investigated using gravitational potential without solving the equations of motion\cite{Shaymatov:2020tna}, but some aspects without exact calculations still remain unclear. In this study, we used {\it fully separated} scalar field equations\cite{Frolov:2008jr,Lunin:2017drx} to prove the validity of WCC conjecture for {\it MP black holes with arbitrary rotations and dimensions without any restriction}. Furthermore, the initial MP black hole was assumed to be changed by the fluxes of scalar field carrying conserved charges. Changes in the MP black hole can be obtained up to the first order by the fluxes during the infinitesimal time interval. Hence, the proof provided in this study is {\it completely valid} for general MP black holes with a massless scalar field of first order. We further investigated the physical implications of our analysis in terms of thermodynamics. We found that the changes in MP black holes correspond to the laws of thermodynamics including the third law. In particular, from a thermodynamic perspective, the changes are specifically directional, which {\it supports} the WCC conjecture. Thus, the WCC conjecture is valid for high-dimensional MP black holes with arbitrary rotations.

The remainder of this paper is organized as follows. In Section \ref{sec:02}, we review MP black holes and the separable Klein-Gordon equation. In Section \ref{sec:03}, we highlight the changes in MP black hole that were obtained from the fluxes of scalar field. In Section \ref{sec:04}, the WCC conjecture is investigated in extremal and near-extremal MP black holes with general rotations. In Section \ref{sec:05}, the changes in MP black hole that are related to the laws of thermodynamics are highlighted. Section \ref{sec:06} summarizes the results obtained.

\section{Separable Klein-Gordon Equation in MP Black Holes}\label{sec:02}

The MP black hole is a solution to Einstein's theory of gravity for higher dimensions\cite{Myers:1986un}. The generalization of the Kerr black hole indicates that it has more than one rotating plane to define the independent angular momenta owing to additional spatial dimensions. Because angular momentum is defined by a pair of spatial dimensions, the even-dimensional black hole has an additional spatial dimension that cannot be included in rotating planes, and the odd-dimensional one has no such dimension. Hence, the metric of the $D$-dimensional MP black hole is given in two forms for the even ($D=2n+2$) and odd ($D=2n+1$) dimensions. The even-dimensional MP black hole is given by
\begin{align}
\text{(even $D$)}\quad ds^2&=-dt^2+\frac{Mr}{FR}\left(dt+\sum_{i=1}^{n}a_i \mu_i^2 d\phi_i\right)^2+\frac{FR}{R-Mr}dr^2+\sum_{i=1}^{n}(r^2+a_i^2)(d\mu_i^2+\mu_i^2d\phi_i^2)+r^2d\alpha^2,\\
F&=1-\sum_{k=1}^{n}\frac{a_k^2 \mu_k^2}{r^2+a_k^2},\quad R=\prod_{k=1}^n(r^2+a_k^2),\nonumber
\end{align}
where $M$ is the mass parameter, and $a_k$ is the $k$th spin parameter. The additional spatial dimension with $(\alpha,\mu_i)$ satisfies that $\alpha^2+\sum_{i=1}^{n}\mu_i^2=1$. The number of possible spin parameters is $n$, which is the same as the number of rotating planes for given dimensions. The odd-dimensional metric is given by
\begin{align}
\text{(odd $D$)}\quad ds^2=-dt^2+\frac{Mr^2}{FR}\left(dt+\sum_{i=1}^{n}a_i \mu_i^2 d\phi_i\right)^2+\frac{FR}{R-Mr^2}dr^2+\sum_{i=1}^{n}(r^2+a_i^2)(d\mu_i^2+\mu_i^2d\phi_i^2),
\end{align}
where $\sum_{i=1}^{n}\mu_i^2=1$. The mass and spin parameters are associated with the mass $M_\text{B}$ and angular moment $J_k$ of the MP black hole. 
\begin{align}
\sigma_D\equiv\frac{\Omega_{D-2}}{8\pi G_D},\quad M_\text{B}=\frac{D-2}{2}\sigma_D M,\quad J_k=\sigma_D M a_k,
\end{align}
where we will set $G_D$ to unity. $\sigma_D$ is a dimensional property. Each angular momentum is treated as an independent variable in a particular direction. Subsequently, the corresponding $n$ independent angular velocities can be defined at the outer horizon $r_\text{h}$. The $k$th angular velocity is denoted as $\Omega^k_\text{h}$.
\begin{align}
\Omega^k_\text{h}=\frac{a_k}{r_\text{h}^2+a_k^2}.
\end{align}
The surface area $A_\text{h}$ is associated with surface gravity $\kappa_\text{h}$\cite{Emparan:2008eg,Gwak:2011rp}.
\begin{align}\label{eq:thermodynamicproperties02}
\quad A_\text{h}&=\frac{\Omega_{D-2}}{2\kappa_\text{h}}M\left(D-3-\sum_i\frac{2a_i^2}{r_\text{h}^2+a_i^2}\right)=\Omega_{D-2}r_\text{h}M,\\
\kappa_\text{h}&=\text{(even $D$)}\lim_{r\rightarrow r_\text{h}} \frac{\partial_r R-M}{2M r}\,\,\text{or}\,\,\text{(odd $D$)}\lim_{r\rightarrow r_\text{h}} \frac{\partial_r R-2M r}{2M r^2}\nonumber\\
\text{(even and odd $D$)}\quad &=\frac{1}{2r_\text{h}}\left(D-3-\sum_k \frac{2a_k^2}{r_\text{h}^2+a_i^2}\right).\nonumber
\end{align}
The Bekenstein-Hawking entropy and Hawking temperature are given as
\begin{align}
S_\text{h}=\frac{1}{4}A_\text{h},\quad T_\text{h}=\frac{\kappa_\text{h}}{2\pi}.
\end{align}

Our investigation was based on the separability of Klein-Gordon equation to MP black holes as mentioned in \cite{Lunin:2017drx}. Hence, we have briefly reviewed the separation of Klein-Gordon equations in arbitrary dimensions. However, the process to obtain the equations is complicated ; therefore, we have summarized it as important steps and identifications that have been used in this study. The details given in \cite{Lunin:2017drx} have been modified to adapt to our notations.

Beginning with Klein-Gordon equation with respect to the massless scalar field,
\begin{align}\label{eq:kleingordoneq01}
\frac{1}{\sqrt{-g}}\partial_\mu\left(\sqrt{-g} g^{\mu} \partial_\nu \Psi\right)=0,
\end{align}
which is separable in elliptic coordinates $x_k$. The elliptical coordinate rewrites the directional cosine as
\begin{align}
\text{(even $D$)}&\quad (a_i\mu_i)^2=\frac{1}{c_i^2}\prod^n_{k=1}(a_i^2-x_k^2),\quad c_i=\prod_{k\not= i}(a_i^2-a_k^2),\quad 0<x_1<a_1<\dotsb<x_n<a_n,\\
\text{(odd $D$)}& \quad \mu_i^2=\frac{1}{c_i^2}\prod^{n-1}_{k=1}(a_i^2-x_k^2).\nonumber
\end{align}
The inverse metric can be constructed using the eigenvectors of Killing-Yano tensors\cite{Chervonyi:2015ima}.
\begin{align}
\text{(even $D$)}\,\,e_t&=-\sqrt{\frac{R^2}{FR(R-Mr)}}\left(\partial_t - \sum_k \frac{a_k}{r^2+a_k^2}\partial_{\phi_k}\right),\,\, e_i=-\sqrt{\frac{H_i}{d_i(r^2+x_i^2)}}\left(\partial_t-\sum_k \frac{a_k}{a_k^2-x_i^2}\partial_{\phi_k}\right),\\
e_r&=\sqrt{\frac{R-Mr}{FR}}\partial_r,\quad e_{x_i}=\sqrt{\frac{H_i}{d_i(r^2+x_i^2)}}\partial_{x_i},\quad R=\prod_k(r^2+a_k^2),\quad FR=\prod_k (r^2+x_k^2),\nonumber\\
d_i&=\prod_{k\not=i}(x_k^2-x_i^2),\quad H_i=\prod_k (a_k^2-x_i^2),\quad G_i=\prod_k (a_i^2-x_k^2),\quad c_i^2=\prod_{k\not= i}(a_i^2-a_k^2),\nonumber\\
\text{(odd $D$)}\,\,e_t&=-\sqrt{\frac{R^2}{FR(R-Mr^2)}}\left(\partial_t - \sum_k \frac{a_k}{r^2+a_k^2}\partial_{\phi_k}\right),\,\, e_i=-\sqrt{\frac{H_i}{x_i^2 d_i(r^2+x_i^2)}}\left(\partial_t-\sum_k \frac{a_k}{a_k^2-x_i^2}\partial_{\phi_k}\right),\nonumber\\
e_r&=\sqrt{\frac{R-Mr^2}{FR}}\partial_r,\quad e_{x_i}=\sqrt{\frac{H_i}{x_i^2 d_i(r^2+x_i^2)}}\partial_{x_i},\quad FR=r^2\prod_k (r^2+x_k^2).\nonumber
\end{align}
Furthermore, the determinant of the metric is given by
\begin{align}
\text{(even $D$)}\,\, \sqrt{-g}=\frac{FR}{\prod a_i} \sqrt{\prod\frac{d_i}{c_i^2}},\quad \text{(odd $D$)}\,\,\sqrt{-g}=\frac{FR}{r}\sqrt{\prod\frac{x_i^2 d_i}{c_i^2}}.
\end{align}
The Klein-Gordon equation is separable under the assumption of an ansatz.
\begin{align}
\Psi = e^{i\omega t +i \sum m_k \phi_k}\Phi(r) \prod_i X_i(x_i).
\end{align} 
The nontrivial parts of the separated equations are approximately $r$ and $x_i$ coordinates.
\begin{align}\label{eq:speq01}
\text{(even $D$)}\quad&\frac{d}{dr}\left((R-Mr)\frac{d\Phi}{dr}\right)+\frac{R^2}{R-Mr}\left(\omega-\sum_k \frac{a_km_k}{r^2+a_k^2}\right)^2\Phi-P_{n-1}(r^2)\Phi=0,\\
&\frac{d}{dx_i}\left(H_i \frac{dX_i}{dx_i}\right)-H_i\left(\omega-\sum_k\frac{a_km_k}{a_k^2-x_i^2}\right)^2 X_i+P_{n-1}(-x_i^2)X_i=0,\nonumber\\
\text{(odd $D$)}\quad&r\frac{d}{dr}\left(\frac{(R-Mr^2)}{r}\frac{d\Phi}{dr}\right)+\frac{R^2}{R-Mr^2}\left(\omega-\sum_k \frac{a_km_k}{r^2+a_k^2}\right)^2\Phi-P_{n}(r^2)\Phi=0,\nonumber\\
&\frac{d}{dx_i}\left(\frac{H_i}{x_i} \frac{dX_i}{dx_i}\right)-H_i\left(\omega-\sum_k\frac{a_km_k}{a_k^2-x_i^2}\right)^2 X_i+P_{n}(-x_i^2)X_i=0,\nonumber
\end{align} 
where $P_{n}$ denotes an arbitrary polynomial with $n$ degrees. Note that the form of $P_n$ polynomials is not significant in our investigation because we focus on the near-horizon geometry where the polynomials are negligible for solving the equations. The details of this are discussed in the following sections.

To match the conventions of Kerr black holes in previous studies, we have modified the signs in $\omega$ and $a_k$ from \cite{Lunin:2017drx}.
\begin{align}
\omega\rightarrow -\omega,\quad a_k\rightarrow -a_k.
\end{align}
However, this does not change the physical meaning, and the choice of these signs correspond to well-known forms of expressions in the solutions to scalar field equations.

\section{Fluxes on Outer Horizon}\label{sec:03}

The state of an MP black hole depends on its conserved charges, such as mass and angular momenta. Here, the scalar field can carry its energy and angular momenta into the black hole; thus, the carried charges change the state of the black hole. However, the variations correspond to the fluxes of energy and angular momenta of the scalar field owing to the conservation.

Beginning with the radial equations in Eqs.\,(\ref{eq:speq01}), we rewrite it to the Schr\"{o}dinger-like equation given by
\begin{align}\label{eq:transf01}
\text{(even $D$)}\quad\Phi\rightarrow \frac{\Phi}{\sqrt{R}},\quad \frac{dr^*}{dr}=\frac{R}{R-Mr},\quad \text{(odd $D$)}\quad\Phi\rightarrow \frac{\Phi}{\sqrt{R/r}},\quad \frac{dr^*}{dr}=\frac{R}{R-Mr^2}.
\end{align}
Note that the radial interval changes to $(-\infty,+\infty)$ in the tortoise coordinate. Then, under Eq.\,(\ref{eq:transf01}), the radial equation becomes Schr\"{o}dinger-type equation.
\begin{align}\label{eq:radialsch01}
\text{(even $D$)}\,\,&\frac{d^2\Phi}{d{r^*}^2}+\frac{R-Mr}{4R^3}\left(-2 R \frac{dR}{dr}\left(\frac{M}{R^2}\left(r\frac{dR}{dr}-R\right)\right)+\left(\frac{R-Mr}{R}\right)\left(\left(\frac{dR}{dr}\right)^2-2R \frac{d^2 R}{dr^2}\right)\right)\Phi\\
&+\left(\omega-\sum_{k} \frac{a_{k} m_{k}}{r^{2}+a_{k}^{2}}\right)^{2}\Phi - \left(\frac{R-Mr}{R^2}\right)P_{n-1}\left(r^{2}\right)\Phi=0.\nonumber\\
\text{(odd $D$)}\,\,&\frac{d^2\Phi}{d{r^*}^2}+\frac{R-Mr^2}{4R^3}\left(\left(-R+r\frac{dR}{dr}\right)\left(M+\frac{3R}{r^2}-\frac{3Mr}{R}\frac{dR}{dr}+\frac{1}{r}\frac{dR}{dr}\right)-2\left(R-Mr^2\right)\frac{d^2R}{dr^2}\right)\Phi\nonumber\\
&+\left(\omega-\sum_k\frac{a_k m_k}{r^2+a_k^2}\right)^2\Phi-\left(\frac{R-Mr^2}{R^2}\right)P_n(r^2)\Phi=0.\nonumber
\end{align}
Based on the radial equation in Eq.\,(\ref{eq:radialsch01}), the energy extracted or absorbed by the black hole can be measured at two boundaries: asymptotic region and outer horizon. The limit $r\rightarrow \infty$ yields the radial equation in Eq.\,(\ref{eq:radialsch01}), which is very simple to solve and is given as
\begin{align}
\text{(even and odd $D$)}\quad\frac{1}{\Phi}\frac{d^2\Phi}{{dr^*}^2}+\omega^{2}=0,
\end{align}
where the solution is
\begin{align}\label{eq:solaymp01}
\Phi(r^*)=\mathcal{O}e^{i\omega r^*}+\mathcal{I}e^{-i\omega r^*}.
\end{align}
Note that $\mathcal{O}$ and $\mathcal{I}$ are the outgoing and ingoing amplitudes, respectively. However, in the limit $r\rightarrow r_\text{h}$, the radial equation in Eq.\,(\ref{eq:radialsch01}) can be rewritten as
\begin{align}
\text{(even and odd $D$)}\quad \frac{1}{\Phi}\frac{d^2\Phi}{{dr^*}^2}+\left(\omega-\sum_{k} m_k\Omega^k_\text{h}\right)^{2}=0,
\end{align}
which has two solutions corresponding to the incoming and outgoing waves from the outer horizon. However, the scalar field allows only an ingoing wave into the outer horizon\cite{Brito:2015oca,Berti:2009kk}. Thus, the physical solution is chosen as
\begin{align}\label{eq:solouter01}
\Phi(r^*)=\mathcal{T}e^{-i(\omega-\sum_{k} m_k\Omega_k)r^*},
\end{align}
where $\mathcal{T}$ denotes the transmission amplitude. The amplitudes of the solutions in Eqs.\,(\ref{eq:solaymp01}) and (\ref{eq:solouter01}) are closely associated with each other. This can be represented in terms of the Wronskian as
\begin{align}
W[\Phi(r^*),\Phi^*(r^*)]=\frac{d\Phi}{dr^*}\Phi^*-\Phi\frac{d\Phi^*}{dr^*},
\end{align}
which should be equal to both the boundaries.
\begin{align}
\lim_{r^*\rightarrow -\infty}W[\Phi(r^*),\Phi^*(r^*)]=\lim_{r^*\rightarrow \infty}W[\Phi(r^*),\Phi^*(r^*)].
\end{align}
It produces the relationship between amplitudes at the boundaries.
\begin{align}
|\mathcal{O}|^2=|\mathcal{I}|^2 - \frac{\omega-\sum_k m_k \Omega_k}{\omega}|\mathcal{T}|^2.
\end{align}
This indicates that the absorption or superradiance depends on the frequency of the scalar field. In addition, the black hole absorbs the scalar field when $|\mathcal{O}|<|\mathcal{I}|$. The frequency satisfies
\begin{align}\label{eq:absorption01}
\omega>\sum_k m_k \Omega_k.
\end{align}
Furthermore, given by $|\mathcal{O}|<|\mathcal{I}|$, the superradiant condition is given by
\begin{align}\label{eq:superrdiance01}
\omega<\sum_k m_k \Omega_k.
\end{align}
Thus, the reaction of the scattering process depends on the states of both the black hole and scalar field.
  
In this study, we obtained the amount of conserved charges carried into the black hole using the scalar field. Based on the conserved charges, the initial black hole can be in a different state represented as $(M,J_k)$. The difference from the initial state to final state exactly coincides with the amount of carried conserved charges, which can be measured by the fluxes of the scalar field at the outer horizon. Beginning with the solution in the outer horizon from Eqs.\,(\ref{eq:solouter01}),
\begin{align}\label{eq:outerfullsol01}
\text{(even $D$)}\,\,\Psi&=\frac{\mathcal{T}}{\sqrt{R_\text{h}}}e^{-i\omega t}e^{-i(\omega-\sum_{k} m_k\Omega^k_\text{h})r^*}e^{i\sum_{k}m_k\phi_k}\prod_i X_i(x_i),\quad R_\text{h}=R\left|_{r=r_\text{h}}\right.,\\
\text{(odd $D$)}\,\,\Psi&=\frac{\mathcal{T}}{\sqrt{R_\text{h}/r_\text{h}}}e^{-i\omega t}e^{-i(\omega-\sum_{k} m_k\Omega^k_\text{h})r^*}e^{i\sum_{k}m_k\phi_k}\prod_i X_i(x_i).\nonumber
\end{align}
The conserved charges are carried by the scalar field and encoded in the energy-momentum tensor. Because the conserved charges transferred into the black hole can be measured at the outer horizon, the energy-momentum tensor must be computed at the outer horizon using Eq.\,(\ref{eq:outerfullsol01}) as
\begin{align}
T_{\nu}^{\mu}=\frac{1}{2} \partial^{\mu} \Psi \partial_{\nu} \Psi^{*}+\frac{1}{2} \partial^{\mu} \Psi^{*} \partial_{\nu} \Psi-\frac{1}{2}\delta_{\nu}^{\mu} \partial_{\rho} \Psi \partial^{\rho} \Psi^{*}.
\end{align}
Then, the amount of energy and angular momenta of the scalar field flowing into the black hole can be counted from the fluxes in the outer horizon. Because the energy and angular momenta should be conserved in spacetime, we can assume that they are still preserved by absorption into the conserved charges of the black hole, mass, and angular momenta even if they cannot be observed outside the black hole. At the outer horizon, the effects of energy and angular momenta are given by
\begin{align}\label{eq:fluxesenergyangular}
\text{(even and odd $D$)}\quad\frac{dE}{dt}&=\int T^r_t \sqrt{-g} \left(\prod dx_i\right)\left(\prod d\phi_i\right)=|T|^2 \chi^2 \omega(\omega-\sum_k m_k\Omega^k_\text{h}),\\
\frac{dL_k}{dt}&=-\int T^r_{\phi_k} \sqrt{-g} \left(\prod dx_i\right)\left(\prod d\phi_i\right)=|T|^2 \chi^2 m_k(\omega-\sum_k  m_k\Omega^k_\text{h}),\nonumber
\end{align}
where
\begin{align}
\text{(even $D$)}&\quad\chi^2=\int \left(\prod dx_i\right)\left(\prod d\phi_i\right) \frac{1}{\left(\prod a_i\right)}\sqrt{\prod \left(\frac{d_i}{c_i^2}\right)}\prod_i X_i(x_i)\prod_j X_j^*(x_j),\\
\text{(odd $D$)}&\quad\chi^2=\int \left(\prod dx_i\right)\left(\prod d\phi_i\right)\sqrt{\prod \left(\frac{x_i^2 d_i}{c_i^2}\right)}\prod_i X_i(x_i)\prod_j X_j^*(x_j).
\nonumber
\end{align}
The integration always produces a positive constant with spin parameters. The forms of fluxes are the same in spacetime dimensions because the differences are included in the constant $\chi$, which is not important in our analysis. Note that this is similar to the integration of hyperspheroidal functions in four dimensions.

The fluxes in Eq.\,(\ref{eq:fluxesenergyangular}) carry energy and angular momenta into the black hole. During an infinitesimal time interval $dt$, the changes in the mass and angular momenta of the black hole should coincide with the fluxes to be preserved.
\begin{align}\label{eq:changesenergyangular}
dM_\text{B}&=|T|^2 \chi^2 \omega(\omega-\sum_k m_k\Omega^k_\text{h})dt,\quad dJ_k=|T|^2 \chi^2 m_k(\omega-\sum_k  m_k\Omega^k_\text{h})dt.
\end{align}
The fluxes of the scalar field can change the mass and angular momenta according to the absorption and superradiance conditions in Eqs.\,(\ref{eq:absorption01}) and (\ref{eq:superrdiance01}), respectively. Here, the state of the black hole depending on mass and angular momenta can change during an infinitesimal time interval. For convenience, the changes in mass and spin parameters can be obtained using Eqs.\,(\ref{eq:changesenergyangular}).
\begin{align}
\text{(even and odd $D$)}\quad
dM&=\frac{2}{(D-2)\sigma_D}|T|^2 \chi^2\omega(\omega-\sum_k m_k\Omega_\text{h}^k )dt,\\
da_k&=\frac{1}{\sigma_DM}\left(\frac{m_k}{\omega}-\frac{2a_k}{D-2}\right)|T|^2 \chi^2\omega(\omega-\sum_k m_k\Omega_\text{h}^k )dt.\nonumber
\end{align}
The state of the black hole is then given by independent variables $(M,a_k)$ rather than $(M_\text{B},J_k)$.

\section{Weak Cosmic Censorship Conjecture}\label{sec:04}

The scalar field carries energy and angular momenta to the black hole resulting in the changes in the mass and angular momenta of the black hole. Furthermore, during the infinitesimal time interval, the initial state of the black hole given by $(M,a_k)$ ends at the final state of $(M+dM,a_k+da_k)$. These changes can affect various properties of the black hole including the horizon locations. In particular, the changes in mass and angular momenta affect the extremal condition that black holes should satisfy according to the WCC conjecture.

\begin{figure}[h]
\centering
\subfigure[{$\Delta$ with $(M,a_1)$ in $D=4$.}] {\includegraphics[scale=0.6,keepaspectratio]{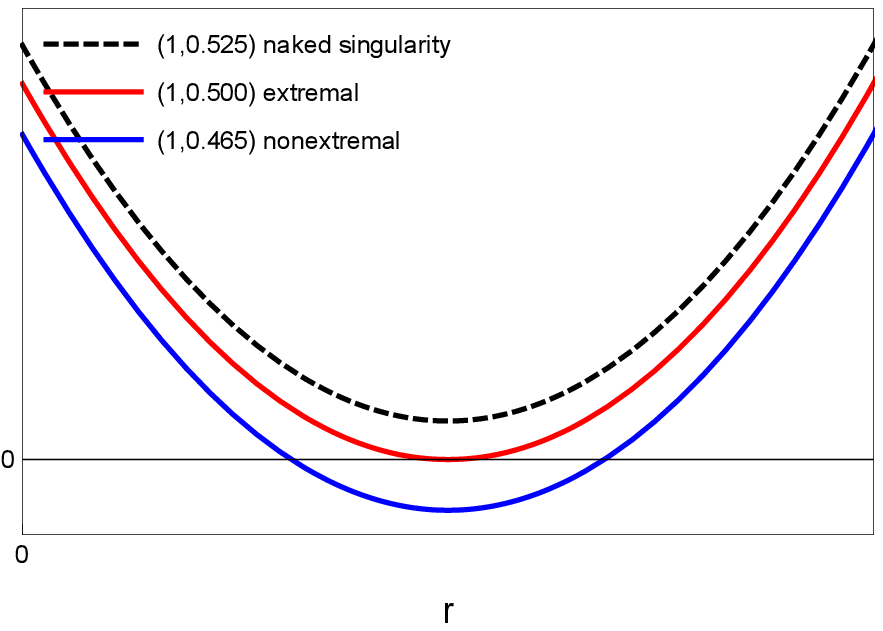}}\quad
\subfigure[{$\Delta$ with $(M,a_1,a_2)$ in $D=6$.}] {\includegraphics[scale=0.6,keepaspectratio]{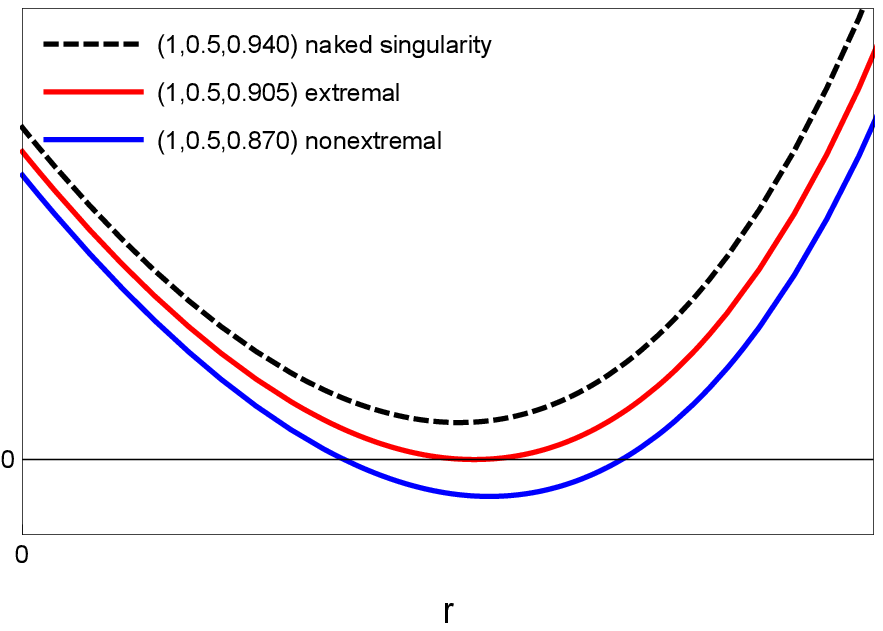}}\quad
\subfigure[{$\Delta$ with $(M,a_1,a_2,a_3)$ in $D=8$.}] {\includegraphics[scale=0.6,keepaspectratio]{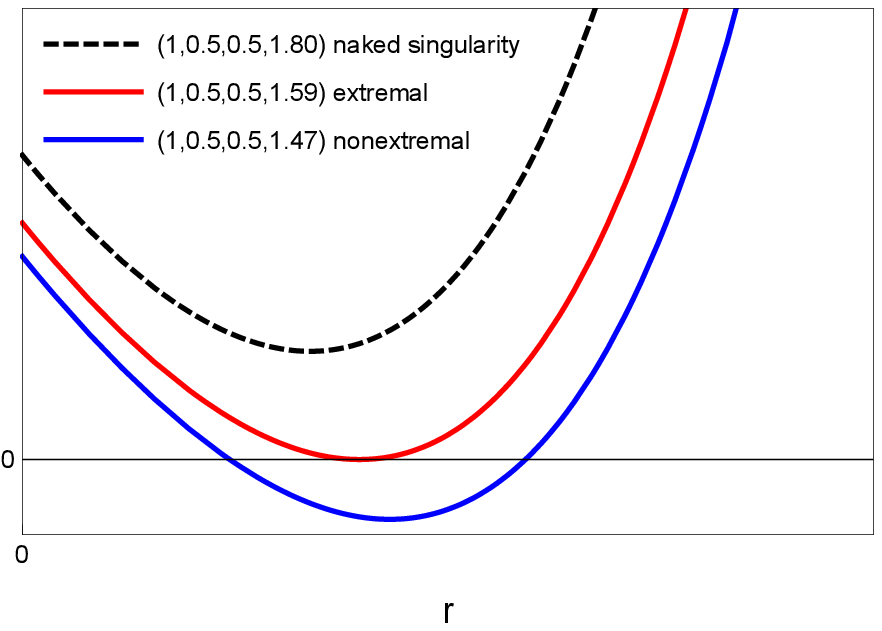}}
\caption{{\small The extremality of MP black holes in various dimensions by $\Delta (M, a_k,r)$.}}
\label{fig:dimensional01}
\end{figure}

Herein, we investigated the WCC conjecture for changes in mass and angular momenta based on the scattering process of the scalar field. These changes can affect the balance between mass and angular momenta for extremal conditions; therefore, the inner and outer horizons can disappear in spacetime. This is directly related to whether the black hole can be overspun beyond extremal conditions into a naked singularity. Then, we examined the metric function that governs the locations of the horizons.
\begin{align}\label{eq:delta02}
\text{(even $D$)}\,\, \Delta (M,a_k,r)\equiv R-Mr,\quad \text{(odd $D$)}\,\, \Delta (M,a_k,r)\equiv R-Mr^2.
\end{align}
To investigate the WCC conjecture, the existence of horizons must be determined by solving Eq.\,(\ref{eq:delta02}). It can be simplified into the sign of the minimum value of $\Delta(M,a_k,r)$ because the minimum value is negative for nonextremal black holes, zero for extremal holes, and positive for naked singularity as shown in Fig.\,\ref{fig:dimensional01}.
Therefore, the final state of the black hole can be determined by the minimum value of $\Delta$ in Eq.\,(\ref{eq:delta02}). The mass and angular momenta of the black hole change according to the fluxes of the scalar field owing to the scattering process. The function $\Delta$ of the variables $(M,a_k,r)$ also depends on mass and angular momenta and changes into $\Delta(M+dM,a_k+da_k,r)$ in the final state. Subsequently, the minimum location and its value move into $r_\text{min}+dr_\text{min}$ and $\Delta(M+dM,a_k+da_k,r_\text{min}+dr_\text{min})$, respectively. For simplicity, we denote that
\begin{align}
\Delta_\text{min}\equiv \left.\Delta(M,a_k,r)\right|_{r=r_\text{min}},\quad \Delta (M+dM,a_k+da_k,r_\text{min}+dr_\text{min})\equiv\Delta_\text{min}+d\Delta_\text{min}.
\end{align}
Then, during the infinitesimal time interval, the change in minimum value is given by
\begin{align}\label{eq:minimumvalue01}
\Delta_\text{min}+d\Delta_\text{min}=\Delta_\text{min}+\frac{\partial \Delta_\text{min}}{\partial M} d M+\sum_k\frac{\partial \Delta_\text{min}}{\partial a_k} d a_k+\frac{\partial \Delta_\text{min}}{\partial r_\text{min}} d r_\text{min}.
\end{align}
According to Eq.\,(\ref{eq:minimumvalue01}), we can determine the value of the minimum for a given initial condition. In our investigation, it is important to determine whether an initial black hole can be a naked singularity based on small changes in its mass and angular momenta. Thus, by saturating the extremal condition, extremal and near-extremal MP black holes with general rotations are the focus of our investigations.

\subsection{Extremal MP Black Holes with General Rotations}\label{sec:041}

The extremal black holes have saturated spin parameters under extremal conditions. Even if the fluxes of the scalar field infinitesimally change the mass and angular momenta, it is possible that the angular momenta will exceed the extremal condition. The initial state of the extremal black hole is given as
\begin{align}\label{eq:extremalstate01}
r_\text{min}=r_\text{h},\quad \Delta_\text{min}=0,\quad \left.\frac{\partial \Delta}{\partial r}\right|_{r=r_\text{min}}=\frac{\partial \Delta_\text{min}}{\partial r_\text{min}}=0,
\end{align}
which implies that the inner horizon, outer horizon, and minimum location are coincident to a point as shown in Fig.\,\ref{fig:dimensional01}. Under the initial state in Eq.\,(\ref{eq:extremalstate01}), the minimum value in Eq.\,(\ref{eq:minimumvalue01}) is given by
\begin{align}
d\Delta_\text{min}=\frac{\partial \Delta_{\min }}{\partial M} d M+\sum_k\frac{\partial \Delta_{\min }}{\partial a_k} d a_k
\end{align}
with
\begin{align}
\text{(even $D$)}\,\,\frac{\partial \Delta_{\min }}{\partial M}=-r_\text{h},\,\, \frac{\partial \Delta_{\min }}{\partial a_k}=2R_\text{h}\Omega_\text{h}^{k},\quad \text{(odd $D$)}\,\,\frac{\partial \Delta_{\min }}{\partial M}=-r_\text{h}^2,\,\, \frac{\partial \Delta_{\min }}{\partial a_k}=2R_\text{h}\Omega_\text{h}^{k},
\end{align}
where $r_\text{min}$ can be exchanged with $r_\text{h}$ under the initial conditions in Eq.\,(\ref{eq:extremalstate01}). Using analytical calculations, we can obtain the minimum value in the final state.
\begin{align}\label{eq:extremalblackholechange15}
\text{(even $D$)}\,\,d\Delta_\text{min}=-\frac{2|T|^2 r_\text{h} \chi^2}{\sigma_D}(\omega-\sum_k m_k\Omega_\text{h}^k )^2dt,\quad \text{(odd $D$)}\,\,d\Delta_\text{min}=-\frac{2|T|^2 r_\text{h}^2 \chi^2}{\sigma_D}(\omega-\sum_k m_k\Omega_\text{h}^k )^2dt.
\end{align}
It is always {\it{negative}} for any mode of the scalar field in even and odd dimensions. Note that it can be zero for zero flux. In terms of Fig.\,\ref{fig:dimensional01}, the initial states of red lines become the final states of blue lines. Therefore, the extremal MP black hole cannot be overspun based on the scattering of the scalar field including superradiance, thus developing into a nonextremal black hole in the final state. This implies that the final set of spin parameters remains lower than the saturated set to the extremal condition for the final mass. Then, the singularity still hides inside the outer horizon and no observer can see it. Thus, the WCC conjecture is valid for extremal MP black holes.

\subsection{Near-Extremal MP Black Holes with General Rotations}\label{sec:042}

Here, we considered a near-extremal black hole as the initial state. The near-extremal cases do not saturate the extremal condition, which differs somewhat from the extremal cases. Furthermore, the minimum value is a very small negative value that might change to a positive value owing to the changes in mass and angular momenta. In particular, they can demonstrate a change in the state of nonextremal black holes.

The initial state for near-extremal black holes is
\begin{align}\label{eq:nearextremalstate01}
r_\text{h}-r_\text{min}=\epsilon\ll 1,\quad -\Delta_\text{min}\ll 1,\quad \frac{\partial \Delta_\text{min}}{\partial r_\text{min}}=0,\quad \left.\Delta\right|_{r=r_\text{h}}=\Delta_\text{h}=0,\quad \left.\frac{\partial \Delta}{\partial r}\right|_{r=r_\text{h}}=\frac{\partial \Delta_\text{h}}{\partial r_\text{h}}\ll 1,
\end{align}
where $\epsilon$ is the difference between the locations of the outer horizon and the minimum and a very small value comparable to an infinitesimal value to set the near-extremal black holes. The change in the minimum value is then obtained from the variation as
\begin{align}\label{eq:nearextremalchange02}
d\Delta_\text{min}=\frac{\partial \Delta_{\min }}{\partial M} d M+\sum_k\frac{\partial \Delta_{\min }}{\partial a_k} d a_k+\frac{\partial \Delta_{\min}}{\partial r_{\min}}d r_{\min},
\end{align}
where the third term on the left-hand side is zero according to the initial state in Eq.\,(\ref{eq:nearextremalstate01}). As the location of the minimum can be rewritten as that of the outer horizon, each part of Eq.\,(\ref{eq:nearextremalchange02}) can be described in terms of the outer horizon. Then, the other terms are
\begin{align}
\text{(even $D$)}&\quad \frac{\partial \Delta_{\min }}{\partial M}=-r_\text{h}+\epsilon,\quad \frac{\partial \Delta_{\min }}{\partial a_k}=2\Omega_\text{h}^k R_\text{h}+\sum_{i\not= k}\frac{4\Omega_\text{h}^k r_\text{h}R_\text{h}}{r_\text{h}^2+a_i^2}\epsilon+\mathcal{O}(\epsilon^2),\\
\text{(odd $D$)}&\quad \frac{\partial \Delta_{\min }}{\partial M}=-r_\text{h}^2+2r_\text{h}\epsilon+\mathcal{O}(\epsilon^2),\quad \frac{\partial \Delta_{\min }}{\partial a_k}=2\Omega_\text{h}^k R_\text{h}+\sum_{i\not= k}\frac{4\Omega_\text{h}^k r_\text{h}R_\text{h}}{r_\text{h}^2+a_i^2}\epsilon+\mathcal{O}(\epsilon^2).
\nonumber
\end{align}
Hence, we can obtain the change in the minimum value.
\begin{align}\label{eq:changeinnearextremal02}
\text{(even $D$)}\\
d\Delta_\text{min}=&-\frac{2|T|^2 r_\text{h} \chi^2}{\sigma_D}(\omega-\sum_k m_k\Omega_\text{h}^k )^2dt\nonumber\\
&+\left(1+\sum_k \sum_{i\not= k}\frac{4\Omega_\text{h}^k r_\text{h}^2}{r_\text{h}^2+a_i^2}\left(\frac{m_k}{\omega}\left(\frac{D-2}{2}-a_k\right)\right)\right)\left(\frac{2}{(D-2)\sigma_D}|T|^2 \chi^2\omega(\omega-\sum_k m_k\Omega_k )\right)dt\epsilon+\mathcal{O}(\epsilon^2),\nonumber\\
\text{(odd $D$)}\nonumber\\
d\Delta_\text{min}=&-\frac{2|T|^2 r_\text{h}^2 \chi^2}{\sigma_D}(\omega-\sum_k m_k\Omega_\text{h}^k )^2dt\nonumber\\
&+\left(2r_\text{h}+\sum_k \sum_{i\not= k}\frac{4\Omega_\text{h}^k r_\text{h}^3}{r_\text{h}^2+a_i^2}\left(\frac{m_k}{\omega}\left(\frac{D-2}{2}-a_k\right)\right)\right)\left(\frac{2}{(D-2)\sigma_D}|T|^2 \chi^2\omega(\omega-\sum_k m_k\Omega_k )\right)dt\epsilon+\mathcal{O}(\epsilon^2).\nonumber
\end{align}
Because this change occurs during the infinitesimal time interval in Eq.\,(\ref{eq:changeinnearextremal02}), the multiple of $dt$ and $\epsilon$ is very small; therefore, the contribution of the second term is negligible compared with the first term. Hence, the negative first term dominates the change. Furthermore, the initial minimum value is negative according to Eq.\,(\ref{eq:nearextremalstate01}). Thus, the minimum value in the final state remains negative during the change. This implies that the final state is a nonextremal black hole, and the initial near-extremal black hole cannot be overspun by the scattering of the scalar field. Therefore, the WCC conjecture is valid.

\section{Thermodynamics}\label{sec:05}

The validity of WCC conjecture is a consequence of the changes in black holes. These changes occur owing to the fluxes of the scalar field, which are analyzed in terms of thermodynamics to clarify their physical meanings. Then, the carried energy and angular momenta perturb the outer horizon, and the change in the horizon can be analyzed in terms of the horizon area. The change in the horizon area is
\begin{align}\label{eq:changearea01}
dA_\text{h}=\frac{\partial A_\text{h}}{\partial M}dM+\sum_k\frac{\partial A_\text{h}}{\partial a_k}da_k+\frac{\partial A_\text{h}}{\partial r_\text{h}}dr_\text{h},
\end{align}
where the change in the outer horizon satisfies
\begin{align}\label{eq:changearea03}
d\Delta_\text{h}=\frac{\partial \Delta_\text{h}}{\partial M} d M+\sum_k\frac{\partial \Delta_\text{h}}{\partial a_k} d a_k+\frac{\partial \Delta_\text{h}}{\partial r_\text{h}}d r_\text{h}=0,\quad \Delta_\text{h}\equiv \left.\Delta\right|_{r=r_\text{h}}=R_\text{h}-Mr_\text{h},
\end{align}
and
\begin{align}
\text{(even $D$)}&\quad \frac{\partial \Delta_\text{h}}{\partial M}=-r_\text{h},\quad \frac{\partial \Delta_\text{h}}{\partial a_k}=2 \Omega_\text{h}^{k}R_\text{h},\quad \frac{\partial \Delta_\text{h}}{\partial r_\text{h}}=M\left(D-3-\sum_k\frac{2a_k^2}{r_\text{h}^2+a_k^2}\right),\\
\text{(odd $D$)}&\quad \frac{\partial \Delta_\text{h}}{\partial M}=-r_\text{h}^2,\quad \frac{\partial \Delta_\text{h}}{\partial a_k}=2 \Omega_\text{h}^{k}R_\text{h},\quad \frac{\partial \Delta_\text{h}}{\partial r_\text{h}}=Mr_\text{h}\left(D-3-\sum_k\frac{2a_k^2}{r_\text{h}^2+a_k^2}\right).\nonumber
\end{align}
In combination with Eq.\,(\ref{eq:changearea01}) and (\ref{eq:changearea03}), we can obtain the change in the horizon area given by
\begin{align}\label{eq:changeinhorizonarea01}
\text{(even and odd $D$)}\quad dA_\text{h}=\frac{\Omega_{D-2}|T|^2 \chi^2(\omega-\sum_k m_k \Omega_\text{h}^k)^2}{\sigma_D\kappa_\text{h}}dt,
\end{align}
which is positive for all initial states. This implies that the horizontal area of the black hole is irreducible. The black hole then evolves into a larger horizon area than any initial state and mode of scalar field. This is the second law of thermodynamics. Furthermore, the change in entropy is larger in the extremal limit $\kappa_\text{h}\rightarrow 0$ than in the nonextremal black holes. In combination with Eqs.\,(\ref{eq:extremalblackholechange15}) and (\ref{eq:changeinhorizonarea01}), the entropy increases when the initial state is close to an extremal black hole. Therefore, near-extremal black holes prefer to be nonextremal.

During the infinitesimal time interval, the fluxes of the scalar field change the mass and angular momenta of the black hole. Additionally, entropy also increases during the process. These changes are associated with each other. Therefore, we can construct their dispersion relation using Eq.\,(\ref{eq:changeinhorizonarea01}), which is given by
\begin{align}
\text{(even and odd $D$)}\quad dM_\text{B}=T_\text{h}dS_\text{h}+\sum_k \Omega_\text{h}^{k} dJ_k.
\end{align}
This is the first law of thermodynamics. Thus, the scalar field changes the black hole under energy conservation during an infinitesimal interval. This ensures that what we obtain is physically correct and consistent with thermodynamics.

According to our investigation of thermodynamics, near-extremal black holes become nonextremal in the final state with an increase in entropy. In terms of the third law of thermodynamics, we found that this change is clearly directional. Here, we analyze the scattering process using the well-known version of the third law in \cite{Bardeen:1973gs}. The version suggests that a zero Hawking temperature cannot be achieved by a finite physical process. The change in Hawking temperature may be demonstrated by a direct variation in Eq.\,(\ref{eq:thermodynamicproperties02}); however, this is quite complicated algebraically. Instead, the change in the Hawking temperature can be obtained by combining Hawking entropy and temperature in Eq.\,(\ref{eq:thermodynamicproperties02}).
\begin{align}
\text{(even and odd $D$)}\quad dT_\text{h}=\frac{T_\text{h}}{S_\text{h}}dS_\text{h}=\frac{\Omega_{D-2}|T|^2 \chi^2(\omega-\sum_k m_k \Omega_\text{h}^k)^2}{4\sigma_D S_\text{h}}dt,
\end{align}
This ensures that the change in temperature is always positive. This also implies that the nonextremal black hole cannot be an extremal black hole with zero temperature owing to scattering according to this third law of thermodynamics. Furthermore, using the second law in Eq.\,(\ref{eq:changeinhorizonarea01}), the increase in temperature is clearly preferred to the change in the black hole, and it is consistent with any arbitrary dimensions.

\section{Summary}\label{sec:06}

In this study, we investigated the WCC conjecture for MP black holes with arbitrary angular momenta based on the scattering of massless scalar field. The MP black hole in higher-dimensional spacetime can rotate in multiple independent directions, which define the angular momenta. Such angular momenta are also associated with metric and field equations. Therefore, it was very difficult to separate the field equation due to which algebraic analysis was impossible. However, based on elliptic coordinates, the scalar field equation was found to be separable and allowed us to analyze the WCC conjecture in arbitrary dimensions and rotating planes. The even and odd dimensions generate different field equations owing to the additional spatial dimension that is not included in the rotating plane; therefore, we had to analyze the even and odd dimensional cases independently.

The scattering of scalar field carries its energy and angular momenta into the MP black hole. The carried energy and angular momenta are transferred into the mass and angular momenta of the black hole to conserve in the spacetime. The changes in mass and angular momenta were obtained from the fluxes of the scalar field measured at the outer horizon. Then, we identified the changes during the infinitesimal time interval in which the equations were of first order. This allowed us to demonstrate the final state of the black hole that back-reacted from a given initial state under scattering.

We then investigated the WCC conjecture for the initial state with extremal and near-extremal black holes. The objective of this investigation was to clarify that the final state can be a black hole or naked singularity when the initial black hole gains mass and angular momenta from the fluxes of the scalar field. The final state can be specified by the metric function $\Delta$ with changes during the infinitesimal time interval. Thus, we inferred that both extremal and near-extremal black holes in any higher dimension cannot be overspun because the final mass and angular momenta are always under extremal conditions indicating that naked singularity cannot appear under the scattering process. Thus, WCC conjecture is {\it valid for MP black holes with arbitrary rotations and dimensions}.

From the thermodynamic perspective, it was observed that the changes in MP black holes were clearly directional. The mass and angular momenta varied according to the fluxes of the scalar field. These changes result in a dispersion relation similar to the first law of thermodynamics. The entropy of MP black hole increases for any mode of scalar field even with superradiance. This is the second law of thermodynamics. In the analysis of the Hawking temperature of an MP black hole, we found that the temperature always increased in any mode of the scalar field. Hence, an extremal black hole of zero Hawking temperature could not be generated under this physical scattering process starting with the initial nonextremal black hole. This also explains the validity of WCC conjecture. Therefore, we observed that the laws of thermodynamics determine the changes in MP black holes and explain the validity of the WCC conjecture under the scattering of scalar field in higher dimensions with arbitrary rotations.

\vspace{10pt} 

\noindent{\bf Acknowledgments}

\noindent This work was supported by the National Research Foundation of Korea (NRF) grant funded by the Ministry of Science and ICT (NRF-2018R1C1B6004349) and the Ministry of Education (NRF-2022R1I1A2063176) and the Dongguk University Research Fund of 2022. BG appreciates APCTP for its hospitality during the topical research program, {\it Multi-Messenger Astrophysics and Gravitation}.

\bibliography{References}
\bibliographystyle{Refsty}
\end{document}